\documentclass[1p]{elsarticle}
\makeatletter
\def\ps@pprintTitle{%
 \let\@oddhead\@empty
 \let\@evenhead\@empty
 \def\@oddfoot{\centerline{\thepage}}%
 \let\@evenfoot\@oddfoot}
\makeatother

\usepackage[utf8]{inputenc}
\usepackage[english]{babel}
\usepackage[babel]{csquotes}
\usepackage{stmaryrd}
\usepackage{amssymb}
\usepackage{amsmath}
\usepackage{amsthm}

\begin{document}

\begin{frontmatter}
\title{Quantum vacuum energy in General Relativity}

\author{Christian Henke}
\ead{henke@math.tu-clausthal.de}

\address{University of Technology at Clausthal, Department of Mathematics,\\ Erzstrasse 1, D-38678 Clausthal-Zellerfeld, Germany}
\date{\today}

\begin{abstract}
The paper deals with the scale discrepancy between the observed vacuum energy in cosmology and the theoretical quantum vacuum energy (cosmological constant problem). 
Here, we demonstrate that Einstein's equation and an analogy to particle physics leads to the first physical justification of the so-called fine-tuning problem. This fine-tuning could be automatically satisfied with the variable cosmological term $\Lambda(a)=\Lambda_0+\Lambda_1 a^{-(4-\epsilon)}$, $0 < \epsilon \ll 1,$ where $a$ is the scale factor.

As a side effect of our solution of the cosmological constant problem, the dynamical part of the cosmological term generates an attractive force and solves the missing mass problem of dark matter.
\end{abstract}


\end{frontmatter}

\section{Introduction}
Direct cosmological experiments could only observe $5\%$ of the total matter-energy content of the universe. The remaining unknown components dark matter and dark energy are classified by their gravitational effects. The first component behaves attractively and constitutes $26\%$ of the matter-energy density, whereas the dark energy has a fraction of $69\%$ and is responsible for an accelerated expansion of the universe \cite{Ade_Planck_2015}.
Since the 1990s, the cosmological constant $\Lambda$ in Einstein's field equation 
\begin{equation*}
R_{\mu \nu}-\frac{1}{2} g_{\mu \nu} R +\Lambda g_{\mu \nu}= \kappa T_{\mu \nu}, \quad \kappa=\frac{8 \pi G}{c^4},
\end{equation*}
has been used as a simple explanation for an expansion of the universe
(see 
\cite{Carroll_TheCosmologicalConstant_1992,
Carroll_TheCosmologicalConstant_2001,
Sahni.Starobinsky_TheCasefora_2000,
Weinberg_TheCosmologicalConstant_1989}
for further details and
\cite{Wald_GeneralRelativity_1984} for notational conventions).
Due to the unknown form of the underlying energy, this anti-gravitational mechanism is called dark energy. From the source point of view the cosmological constant can be written as an energy-momentum tensor
\begin{equation*}
T^{\Lambda}_{\mu \nu}=\left(\rho_\Lambda + \frac{p_\Lambda}{c^2}\right) u_\mu u_\nu + p_\Lambda g_{\mu \nu} , \quad T^{\text{eff}}_{\mu \nu}=T_{\mu \nu}+T^{\Lambda}_{\mu \nu},
\end{equation*}
where $p_\Lambda=-\Lambda/\kappa, \rho_\Lambda=-p_{\Lambda}/c^2$ and $u_\mu$ is the fluid four-velocity.
Hence, in the absence of matter the cosmological constant could be interpreted as the energy density of the vacuum.
In contrast to the notion of an empty space, 
the quantum field theory defines the vacuum as the state of lowest energy density. 
A comparison of both concepts by
cosmological observations of $\Lambda$ and theoretical calculations of the quantum energy density uncovers a large discrepancy
\begin{equation*}
\frac{\rho_{\Lambda}}{\rho_{\text{vac}}} \approx 10^{-121},
\end{equation*}
which 
is ``the most striking problem in the contemporary fundamental physics'' \cite{Dolgov_ProblemVacuumEnergy_1997}.

Changing the point of view and defining the cosmological constant by $\Lambda=\kappa c^2 \rho,$ where $\rho$ is a density field which contributes to the vacuum energy density, it is natural to assume $\Lambda$ as a dynamical quantity. 
Furthermore, there are no a priori reasons why $\Lambda$ should not vary - as long as the energy conservation 
\begin{equation*}
\nabla^\mu T^\text{eff}_{\mu \nu}=0,
\end{equation*}
is satisfied (see \cite[p. 2]{Overduin_Evolutionofthe_1998}).

The remainder of the paper is organised as follows. In section 2 we review the pressure-free Friedmann equations where the cosmological term $\Lambda$ is a function of the scale factor $a$. Section 3 is devoted to a Klein-Gordon equation for the scale factor.
In order to identify the arrived equation as an Euler-Lagrange equation, the Klein-Gordon equation is transformed to a scale factor independent space-time metric.
Then, the underlying Lagrangian density could be determined and the related energy-momentum tensor of an empty Friedmann universe is studied in section 4. This enabled us to establish a canonical decomposition of the total energy into a cosmological and a remainder term.
Consequently, the fine-tuning of the energy densities is justified. Moreover, we demonstrate with an application of the model $\Lambda(a)=\Lambda_0+\Lambda_1 a^{-(4-\epsilon)}$, $0 < \epsilon \ll 1$ that the total energy density of an empty Friedmann universe equals the quantum zero-point energy. Finally, we show that our solution of the cosmological constant problem explains cosmological observations without the missing mass of dark matter.

\section{A time-dependent cosmological term}
Let $M$ be a 4-dimensional manifold equipped with a metric 
$\bar{g}_{\mu \nu},$
which determines the space-time interval as follows: 
\begin{equation*}
d\bar{s}^2=\bar{g}_{\mu \nu} dx^\mu dx^\nu=-c^2 dt^2+a(t)^2 \left(\frac{1}{1-kr^2} dr^2+r^2 d \Omega^2 \right),
\end{equation*}
where  
$\quad d \Omega^2=d\theta^2 + \sin^2 \theta d \phi^2$ and 
$k$ denotes the curvature parameter of unit $\text{length}^{-2}.$ 
Inserting the metric $\bar{g}_{\mu \nu}$
in Einstein's field equation,
we get Friedmann's equations for the scaling factor $a(t)$
\begin{align}
\frac{\dot{a}^2}{a^2} -  \frac{1}{3} \Lambda +\frac{k}{a^2}&= \frac{\kappa c^2}{3}  \rho,
\label{eq:friedmann0}\\
3\frac{ \ddot{a}}{a} - \Lambda  &=-\frac{\kappa}{2} \left( \rho c^2 + 3p \right).
\label{eq:friedmann1}
\end{align}
Until now, several decay laws in which $\Lambda$ is a function of time, the scale factor, the Hubble Parameter and the deceleration parameter have been discussed in the literature (see \cite{Overduin_Evolutionofthe_1998} for an overview).  
Here, we consider a cosmological term of the form
\begin{equation}
\Lambda=\Lambda(a(t)),
\label{eq:def_lambda}
\end{equation}
It is convenient to include the $\Lambda$-term in the energy-momentum tensor of the right-hand side. Therefore, we define an effective density and pressure field
\begin{equation*}
\rho_{\text{eff}}=\rho + \rho_b, \quad p_{\text{eff}}=p+p_b,
\label{eq:eff_rho_p}
\end{equation*}
where $\rho_b=\Lambda(a)/\kappa c^2$ and $p_b=-\Lambda(a)/\kappa$ denote the background fields. 
Now, Einstein's field equation can be written as
\begin{equation}
\bar{R}_{\mu \nu} =\kappa \left( \bar{T}^{\text{eff}}_{\mu \nu} -\frac{\bar{T}^{\text{eff}}}{2} \bar{g}_{\mu \nu} \right),
\label{eq:einstein2}
\end{equation}
where 
\begin{equation*}
\bar{T}^{\text{eff}}_{\mu \nu}= 
\left(\rho_{\text{eff}}+ \frac{p_{\text{eff}}}{c^2} \right) u_\mu u_\nu +p_{\text{eff}} \bar{g}_{\mu \nu},
\label{eq:eff_energy_momentum_tensor}
\end{equation*}
is the effective energy-momentum tensor. The theory is well-defined if 
\begin{equation*}
\nabla^\mu \bar{T}^\text{eff}_{\mu \nu}=0,
\label{eq:divT*}
\end{equation*}
is satisfied. Moreover, a straightforward calculation gives
\begin{equation}
\frac{d}{da} \left(\rho_{\text{eff}} a^3 \right) +3 \frac{p_{\text{eff}}}{c^2} a^2=0, \quad p'_\text{eff}=0,
\label{eq:divT_rho_p}
\end{equation}
or equivalent
\begin{equation}
\frac{d}{da} \left(\rho a^3 \right) +3 \frac{p}{c^2} a^2=-a^3 \frac{d}{da} \rho_b, \quad p'_\text{eff}=0.
\label{eq:divT_rho_p2}
\end{equation}
Equation $(\ref{eq:divT_rho_p})$ is solved in the case of a matter dominated universe by $p=0$ and 
\begin{equation}
\rho=\frac{F}{a^3}+3 \frac{\int^a \Lambda(\alpha) \alpha^2 \, d\alpha}{\kappa c^2 a^3}-\frac{\Lambda(a)}{\kappa c^2}.
\label{eq:rho}
\end{equation} 
Therefore, it follows 
\begin{align}
\frac{\dot{a}^2}{a^2} - \frac{\int^a \Lambda(\alpha) \alpha^2 \, d\alpha}{a^3} +\frac{k}{a^2}&= \frac{\kappa c^2}{3}  \frac{F}{a^3} ,
\label{eq:friedmann4}\\
3\frac{ \ddot{a}}{a} -\frac{3}{2} \Lambda(a)+ \frac{3}{2} \frac{\int^a \Lambda(\alpha) \alpha^2 \, d\alpha}{a^3}&=-\frac{\kappa c^2}{2} \frac{F}{a^3}.
\label{eq:friedmann5}
\end{align}

\section{Conformally-related trace equation}
In this section we provide the ground for a total energy discussion of an empty Friedmann universe which is compatible with quantum field theory. To do so, we have to establish the notion of total energy on a Lagrangian density such that the Euler-Lagrange equation is consistent with Friedmann's equations. Therefore, we have to find a space-time metric which is independent of the scale factor. In order to do so, we start from the field equation for the Robertson-Walker metric $\bar{g}_{\mu \nu}$ and consider the transformed equation for the conformally-related metric $g_{\mu \nu}.$ 

Let $u$ be a strictly positive $C^\infty(M)-$function. The metric $g_{\mu \nu}=u^{-2} \bar{g}_{\mu \nu}$ is said to be conformally-related to $\bar{g}_{\mu \nu}.$ Introducing the notation $\nabla_{\mu}$ for the covariant derivative, $\Delta=g^{\mu \nu} \nabla_{\mu} \nabla_{\nu}$ and $|\nabla u|^2=g^{\mu \nu} \nabla_{\mu} u\nabla_{\nu}u,$ we note down the relation 
for the Ricci scalar (see \cite[p. 446]{Wald_GeneralRelativity_1984}) 
\begin{equation*}
u^4 \bar{R}
=u^2 R 
-6 u \Delta u.
\end{equation*}

Inserting the relation for the Ricci scalar into the trace of equation $(\ref{eq:einstein2}),$ we see that 
\begin{equation*}
-6 u \Delta u + R u^2 = - \kappa u^4 \bar{T}^{\text{eff}},
\end{equation*}
which leads to 
\begin{equation*}
\Delta u -\frac{R}{6} u +\frac{2}{3}\Lambda(a) u^3= -\frac{\kappa}{6}\left(\rho c^2 -3 p\right) u^3.
\end{equation*}
Up to now the above considerations are valid for every strictly positive $u \in C^\infty(M).$ Fixing this function by $u=a$, 
we get $R=6k$ and a Klein-Gordon equation which allows constant coefficients
\begin{equation*}
\Delta a -k a + \frac{2}{3}\Lambda(a) a^3=- \frac{\kappa}{6}\left(\rho c^2 -3 p\right) a^3.
\label{eq:KGa}
\end{equation*}
By using the conformal time $d\tau=dt/a(t),$ 
the conformally-related metric $g_{\mu \nu}$ 
transforms to a curved Minkowski metric
\begin{equation*}
ds^2=\eta_{\mu \nu} dx^\mu dx^\nu=-c^2 d\tau^2+\frac{1}{1-kr^2} dr^2+r^2 d \Omega^2,
\end{equation*}
and the Klein-Gordon equation could be written in the form 
\begin{equation}
-\ddot{a}(\tau)- \frac{d V(a)}{da}= -\frac{\kappa}{6}\left(\rho c^2 -3 p\right) a^3,\quad V(a)=\frac{1}{2} k a^2  -\frac{2}{3} \int^a \Lambda(\alpha) \alpha^3 \, d\alpha +V_0.
\label{eq:KGb}
\end{equation}
Note that, by setting $\Lambda(a),$ one can realise the usual potentials from scalar field theory.

\section{Vacuum energy}
In this section we investigate the total energy density of an energy-momentum tensor which is generated by the Lagrangian density of an empty Friedmann universe.
According to \cite{Zeldovich_CosmologicalConstantTheory_1968} and \cite{Weinberg_TheCosmologicalConstant_1989}, Lorentz invariance implies that the energy density of the vacuum (zero-point energy) acts like a cosmological constant. Hence, there has to be a decomposition of the form 
\begin{equation}
 \rho_{\Lambda}= \rho_{\text{zpe}}+ \rho_{\text{new}},
\label{eq:finetuning}
\end{equation}
where the total zero-point energy density is defined as the lowest energy (ground state) and can be expressed by 
\begin{equation*}
\rho_{\text{zpe}}
=\frac{\hbar k_{\max}^4}{16 \pi^2 c}
=\frac{c^2}{16 \pi^2 G l_p^2}
=\frac{1}{2\pi \kappa l_p^2 c^2},
\end{equation*} 
(cf. \cite{Carroll_TheCosmologicalConstant_1992} and
\cite{Weinberg_TheCosmologicalConstant_1989}). 
Here, $k_{\max} =1/l_p$ denotes a cut-off wave-number and $l_p$ is the Planck length. 
Using the observed value for $\rho_{\Lambda}$ (cf. \cite{Ade_Planck_2015}), this requires a fine-tuning of 121 orders of magnitude. Until now, no physical justification for equation $(\ref{eq:finetuning})$ is known.  


Since $\eta_{\mu \nu}$ is independent of $a$,
equation $(\ref{eq:KGb})$ is the Euler-Lagrange equation of the Lagrangian
\begin{equation}
L=\frac{1}{\kappa} \left(\frac{1}{2} \dot{a}(\tau)^2-V(a) \right),
\label{eq:lagrangian}
\end{equation} 
and\begin{equation}
T^a_{\mu \nu}=\frac{1}{\kappa}\partial_\mu a(\tau) \partial_\nu a(\tau)+g_{\mu \nu}L,
\label{}
\end{equation}
is the related energy-momentum tensor (cf. \cite[p. 65]{Hawking.Ellis_Thelargescale_1973}). Interpreting the energy-momentum tensor for a perfect cosmic fluid,
we get the total energy density and pressure
\begin{equation}
c^2 \rho_a=T_{00}^a=\frac{1}{\kappa} \left( \frac{1}{2}\dot{a}(\tau)^2+V(a)\right), \quad
p_a=L=\frac{1}{\kappa}\left( \frac{1}{2} \dot{a}(\tau)^2-V(a) \right).
\end{equation}
Further $\dot{a}(\tau)$ depends on the coefficients of $V(a).$ Using equation $(\ref{eq:friedmann0}),$ we get 
\begin{equation*}
\dot{a}(\tau)^2= - k a^2 + \frac{1}{3} \Lambda(a) a^4.
\end{equation*}  
Therefore, the total energy density yields
\begin{equation}
c^2 \rho_a =\frac{1}{\kappa} \left( \frac{1}{6} \Lambda(a) a^4 - \frac{2}{3} \int^a \Lambda(\alpha) \alpha^3 \, d\alpha + V_0 \right).
\label{eq:total_energy}
\end{equation}

Using the setting $V_0=0,$ equation $(\ref{eq:total_energy})$ is the physical justification of equation $(\ref{eq:finetuning})$ and yields 
\begin{equation}
\rho_{\Lambda}(a)=\frac{\Lambda(a)}{\kappa c^2}, \quad
\rho_{\text{zpe},\Lambda}(a)=\frac{6 \rho_a}{a^4}, \quad
\rho_{\text{new},\Lambda}(a)=\frac{4 \int^a \Lambda(\alpha) \alpha^3 \, d\alpha}{\kappa c^2 a^4}.
\label{eq:rho_auto_fine}
\end{equation}
Hence, the fine-tuning happens automatically if an appropriate $\Lambda(a)$ can be found that identifies the total energy of the vacuum in quantum field theory and general relativity. Therefore, we identify the zero-point energy of a quantum system with the total energy density of an empty Friedmann universe. 
In order to do that, we consider the model
\begin{equation}
\Lambda(a)=\Lambda_0 + \Lambda_1 a^{-r},\quad r>0,
\label{eq:lambda_model}
\end{equation}
which was also investigated in \cite{Matyjasek_Cosmologicalmodels_1995}.
Using equations $(\ref{eq:total_energy})$ and $(\ref{eq:rho_auto_fine}),$ it follows that the total energy density is independent of $\Lambda_0,$ i.e. 
\begin{equation*}
\rho_{\text{zpe},\Lambda}=\frac{\Lambda_1}{\kappa c^2} \frac{r}{r-4} a^{-r}, \quad r \neq 4.
\end{equation*}
Identifying the total energy densities and using the convention $a_0=a(t_0)=1$ for today's scale factor, we get
\begin{equation}
\Lambda_1=\frac{r-4}{2 \pi r l_p^2}.
\label{eq:lam1}
\end{equation}
If one can fit $(\ref{eq:lambda_model})$ with cosmological observations ($\Lambda(a)c^2/3H_0^2$ is of order unity), then $4 \int^a \Lambda(\alpha) \alpha^3 \, d\alpha / \kappa c^2 a^4$ is identified as the cancellation mechanism which cancels $121$ decimal places.
As the consequence, there is no scale discrepancy between the total energy densities and the cosmological constant problem is solved.

It remains to consider some observational constraints. 
The dynamics of the universe are determined by the equations $(\ref{eq:rho})$ and $(\ref{eq:friedmann4}),$ which depend on the term
\begin{equation}
\frac{\int^a \Lambda(\alpha) \alpha^2 \, d\alpha}{a^3} =\frac{\Lambda_0}{3} +\frac{\Lambda_1}{3-r}a^{-r}. 
\label{eq:int_lam_a2}
\end{equation} 
Therefore, we get for the matter density and $r \neq 3$
\begin{equation}
\rho=\frac{F}{a^3}+\frac{\Lambda_1}{\kappa c^2} \frac{r}{3-r} a^{-r},
\label{eq:ord_matter}
\end{equation}
which is always positive if $\Lambda_1/(3-r)>0.$ 
Moreover, we assume an initial singularity. 
From equation $(\ref{eq:friedmann4})$ and $(\ref{eq:int_lam_a2})$, we can see that $a$ has an initial singularity if $\Lambda_1/(3-r)>0$ is satisfied again.

It remains to discuss some choices for the parameter $r.$ First, let $0 <r <3.$ Since $(\ref{eq:lam1})$ leads to $\Lambda_1/(3-r)<0,$ we neglect this case. Further we have $3<r<4,$ which gives the compatiblility condition $\Lambda_1/(3-r)>0.$ 
In \cite{Matyjasek_Cosmologicalmodels_1995} and \cite{Kimura.Hashimoto.ea_Effectsonthe_2001}, this case was excluded from the considerations because it was assumed without further substantiation that $\Lambda_1$ is always positive. 
In order to analyse the acceleration behavior, we have to discuss the term 
\begin{equation}
\frac{3}{2} \Lambda(a)- \frac{3}{2} \frac{\int^a \Lambda(\alpha) \alpha^2 \, d\alpha}{a^3}
=\Lambda_0 + \frac{3}{2} \Lambda_1 a^{-r} \frac{r-2}{r-3},
\label{eq:accel_behav}
\end{equation} 
from equation $(\ref{eq:friedmann5}).$
Obviously, it follows from the considered parameter range that the dynamical part of $(\ref{eq:accel_behav})$ has a different sign than the usual cosmological constant term. Consequently, the solution of the cosmological constant problem yields a cosmological term with the attractive effect of dark matter. 


Finally, using the settings
\begin{equation}
\begin{aligned}
\Omega_k&=&-\frac{k c^2}{a_0^2 H_0^2}&,& 
\Omega_m&=& \frac{\kappa c^4}{3} \frac{F}{a_0^3 H_0^2},\\
\Omega_{\Lambda}&=&\frac{\Lambda_0 c^2}{3 H_0^2} &,& \Omega_{dm}&=&\frac{\Lambda_1}{3-r} \frac{c^2}{a_0^r H_0^2},
\end{aligned}
\label{eq:def_omega}
\end{equation}
where $H_0$ denotes the actual Hubble constant,
equation  $(\ref{eq:friedmann4})$ could be written as
\begin{equation*}
1=\Omega_k+\Omega_m+\Omega_{\Lambda}+\Omega_{dm}.
\label{}
\end{equation*}
To relate the last equation with observations (cf. \cite{Ade_Planck_2015}), we consider $(\Omega_k,\Omega_m,\Omega_{\Lambda},\Omega_{dm})=(0,0.05,0.69,0.26)$ and $H_0=67.74 \frac{km}{s\, Mpc}.$ 
Using $r=4-\epsilon,$ it follows from $(\ref{eq:lam1})$ and $(\ref{eq:def_omega})$  that the compatibility condition is fulfilled by 
\begin{equation*}
2.288 \cdot 10^{-122}
=\frac{2 \pi l_p^2 H_0^2 \Omega_{dm}}{c^2}
=\frac{r-4}{r(3-r)}
=\frac{\epsilon}{(4-\epsilon)(1-\epsilon)}
\approx \frac{\epsilon}{4}
\end{equation*}
which is satisfied by $\epsilon=9.151 \cdot 10^{-122}.$
Moreover, for the remaining parameters $\Lambda_0$ and $\Lambda_1$ we get
\begin{equation*}
\Lambda_0=1.110 \cdot 10^{-52},\quad \Lambda_1=-1.394 \cdot 10^{-53}.
\end{equation*}

\section{Concluding remarks}
In this paper, the solution of the cosmological constant problem is demonstrated by an application of a variable cosmological term $\Lambda(a)=\Lambda_0 + \Lambda_1 a^{-(4-\epsilon)}$. 
It has been shown that the expansion field $a$ satisfies the Klein-Gordon equation in a non-dynamical space-time and that a variational cosmological term can realise the usual potentials from particle physics. 
Further it was confirmed that the total energy density of an empty Friedmann universe is related to the cosmological term such that the fine-tuning problem was avoided by the setting $\Lambda_1=- \epsilon/2 \pi (4-\epsilon) l_p^2.$
As a consequence of the constraint $0 < \epsilon <1,$ the initial singularity is guaranteed and the dynamical part of the cosmological term generates the attractive force of dark matter.
Finally, the setting of $\epsilon=9.151 \cdot 10^{-122}$ generates the missing mass of dark matter which constitutes $26\%$ of the matter-energy density.


\end{document}